\documentclass[table]{ccjnl}
\pdfoutput=1
\usepackage{lipsum,amsmath}
\usepackage{cuted}
\usepackage{bm}
\usepackage{citesort}
\DeclareMathOperator{\tr}{Tr}
\newcommand{\ie}{\textit{i}.\textit{e}.}

\graphicspath{{figures/}}

\title{Statistical CSI Based Beamforming for Reconfigurable Intelligent Surface Aided MISO Systems with Channel Correlation}

\author{Haochen Li\inst{1,2}, Pan Zhiwen\inst{1,2,*}, Wang Bin\inst{3,*}, Liu Nan\inst{1}, You Xiaohu\inst{1,2}\corinfo{pzw@seu.edu.cn, 869511890@qq.com}}

\receiveddate{}
\reviseddate{}
\Editor{}

\address[1]{National Mobile Communications Research Laboratory (Southeast University), Nanjing 210096, China}
\address[2]{Purple Mountain Laboratories, Nanjing 211100, China}
\address[3]{Science and Technology on Communication Networks Laboratory, Hebei 050081, China}

\begin{document}

\maketitle

\begin{abstract}
Reconfigurable intelligent surface (RIS) is a promising candidate technology of the upcoming Sixth Generation (6G) communication system for its ability to provide unprecedented spectral and energy efficiency increment through passive beamforming. However, it is challenging to obtain instantaneous channel state information (I-CSI) for RIS, which obliges us to use statistical channel state information (S-CSI) to achieve passive beamforming. In this paper, RIS-aided multiple-input single-output (MISO) multi-user downlink communication system with correlated channels is investigated. Then, we formulate the problem of joint beamforming design at the AP and RIS to maximize the sum ergodic spectral efficiency (ESE) of all users to improve the network capacity. Since it is too hard to compute sum ESE, an ESE approximation is adopted to reformulate the problem into a more tractable form. Then, we present two joint beamforming algorithms, namely the singular value decomposition-gradient descent (SVD-GD) algorithm and the fractional programming-gradient descent (FP-GD) algorithm. Simulation results show the effectiveness of our proposed algorithms and validate that $2$-bits quantizer is enough for RIS phase shifts implementation.
\end{abstract}
\section{Introduction}
\label{Introduction}
\quad To cope with the increasing demand for high data rate and ubiquitous coverage for future wireless communication systems, reconfigurable intelligent surface (RIS), which comprises a large number of low-cost passive reflection elements, has received more and more attention \cite{wu2019towards,ref2,ref3}. Via shaping the reflection of impinging signals with adjustable reflection elements (also known as passive beamforming), RIS is capable of actively controlling the radio propagation environment \cite{ref4,ref5}.

\subsection{Prior Works}
\quad Beamforming for RIS under various setups attracts much research interest \cite{ref6,ref7,ref8,ref9,ref10}. To jointly optimize beamforming matrices of the access point (AP) and RIS, alternate optimization (AO) method together with semi-definite relaxation (SDR) technique is used to reach a locally optimal solution for AP transmission power minimization problem in \cite{ref6}. In a scenario similar to \cite{ref6}, authors in \cite{ref7} also employ AO method to jointly optimize beamforming matrices of AP and RIS. Particularly, fixed point iteration and manifold optimization (MO) method are utilized in each iteration to reach stationary point for received signal power maximization problem. Furthermore, in RIS-aided multiple-input single-output (MISO) system, globally optimal beamformers at AP and RIS that maximize user spectral efficiency (SE) are derived in \cite{ref8} using branch and bound (BnB) method. Capitalizing upon burgeoning deep learning technique, \cite{ref9} proposes to design the RIS passive beamforming matrix via a neural network with unsupervised learning approach. Moreover, authors in \cite{ref10, huang2021multi} exploit deep learning to deal with a joint beamforming problem and enable the designed system to gradually learn from the propagation environment.

Above mentioned works mainly start from the premise that instantaneous channel state information (I-CSI) is known at the AP and RIS. In a RIS-aided system, however, it is challenging to obtain the perfect CSI due to the passive nature and an enormous number of RIS reflection elements \cite{wei2021channel,wei2022joint}. An alternative way is to exploit statistical channel state information (S-CSI) since it is easier to obtain and varies more slowly. {Existing works exploiting S-CSI \cite{ref11,ref12,ref13,ref15,peng2021analysis} assume the involved stochastic channel matrices/vectors have uncorrelated elements.} However, for hardware design consideration, RIS is envisioned to be implemented with small element separation and adopts planar array to contain more reflection elements for better system performance \cite{ref18,ref19}, thus leading the channels between RIS and users to be correlated \cite{ref20}.

{Based on S-CSI that contains channel correlation information, authors in \cite{ref21} study energy efficiency and spectral efficiency tradeoff in RIS-aided MIMO uplink. Alternating algorithms based on random matrix theory (RMT) tools are adopted in~\cite{xu2021sum,9066923,ref22} for jointly calculating AP and RIS beamforming matrices. However, works~\cite{ref21,xu2021sum} only discuss the uplink transmission and \cite{9066923} focuses on the maximization of the minimum (Max-Min) signal-to-interference-plus-noise ratio (SINR). Result in \cite{ref22} is only a large system approximation that describes the system ergodic spectral efficiency (ESE) in large system regime.}

\subsection{Our Contributions}
\quad {Against this background, a RIS-aided MISO multi-user downlink communication system with correlated channels is investigated in this paper. A sum ESE maximization problem for beamforming design based on S-CSI is formulated.}\\
\indent The contributions of this paper are summarized as follows:
\begin{enumerate}
\item In RIS-aided systems, it is challenging to obtain perfect CSI for high-dimensional channels introduced by RIS. To tackle this issue, we propose joint beamforming algorithms based on S-CSI of RIS-user channels and AP-RIS channel. Besides, Adopting S-CSI can also reduce the CSI required by beamforming from the long-term perspective.
\item In this paper, we propose two active beamforming algorithms, namely the singular value decomposition (SVD) based beamforming algorithm, and the fractional programming (FP) based beamforming algorithm, and one active beamforming algorithm, namely the gradient descent (GD) algorithm. The SVD-GD algorithm takes the SVD active beamforming method and the GD passive beamforming method, while the FP-GD algorithm takes the FP active beamforming method and the GD passive beamforming method. 

\item Simulation results are presented to demonstrate effectiveness of our proposed algorithms. {In particular, with S-CSI, the proposed algorithms leads to little performance loss but reduced computational and signaling burden compared with benchmark algorithms. The proposed FP-GD algorithm achieves better performance than the proposed SVD-GD algorithm, while the latter algorithm does not require alternative optimization. Besides, it is worth noting that 2-bits quantizer is enough for RIS phase shifts implementation, which offers useful insights for RIS hardware design.}
\end{enumerate}
\subsection{Organization and Notation}

\quad\textit{Organization}: The remainder of this paper is organized as follows: Section II presents the system model of RIS-aided multi-user MISO downlink system and the problem formulation of sum ESE maximization for beamforming design. In Section III, the SVD based active beamforming and the FP based beamforming algorithm are proposed. In Section IV, the GD algorithm for passive beamforming is proposed. Section V presents the overall algorithms and the required pilot overhead of the proposed algorithms. Section VI shows simulation results and analysis. Finally, conclusions are drawn in Section VII.

\textit{Notations}: Throughout the paper, lowercase letters, lowercase bold letters and capital bold letters (e.g., $a$, $\mathbf{a}$ and $\mathbf{A}$) denote scalars, vectors and matrices, respectively. $\mathbb{C}^{M \times K}$ denotes the $M \times K$ dimensional complex vector space. $(\cdot)^\mathrm{T}$, $(\cdot)^\mathrm{*}$ and  $(\cdot)^\mathrm{H}$ represent the operations of transpose, conjugate and conjugate transpose, respectively. The symbol $ \odot $ denotes the Hadamard product. diag$\left( \mathbf{a} \right)$ denotes the diagonal matrix with $\mathbf{a}$ along its main diagonal. max$(\mathbf{a})$ denotes the index of the largest element of $\mathbf{a}$. $\lvert \cdot \rvert$, $\lVert \cdot \rVert_2$ and $\lVert \cdot \rVert_\mathrm{F}$ represent the modulus, the Euclidean norm and the Frobenius norm, respectively. $[\mathbf{A}]_i$ is the $i$th column of matrix $\mathbf{A}$. $[\mathbf{A}]_{mn}$ denotes the $(m,n)$th element of matrix $\mathbf{A}$. The distribution of a circularly symmetric complex Gaussian (CSCG) random vector with mean vector $\mathbf{a}$ and covariance matrix $\mathbf{A}$ is denoted by $\mathcal{C N}\left(\mathbf{a}, \mathbf{A}\right)$. $\lfloor a\rfloor$ denotes the largest integer that is not greater than $a$.$\mod(a,b)$ denotes the remainder of the Euclidean division of $a$ by $b$. arg$(\cdot)$ means the extraction of phase information. $\mathbf{I}$ denotes an identity matrix with appropriate dimensions. $\mathbb{E}\left\{\cdot\right\}$ represents the statistical expectation operator. Calligraphic letters represent sets, e.g., $\mathcal{A,B}$. The set difference is defined as $\mathcal{A} / \mathcal{B} =\{a \mid a \in \mathcal{A}, a \notin \mathcal{B}\}$.

\section{System Model and Problem Formulation}
\label{OFDM}

\quad In this section, we describe system model and channel model of considered RIS-aided multi-user MISO system first and then formulate sum ESE maximization problem for beamforming design.

\begin{figure}[htbp]
\centering
\includegraphics[width=3.5in]{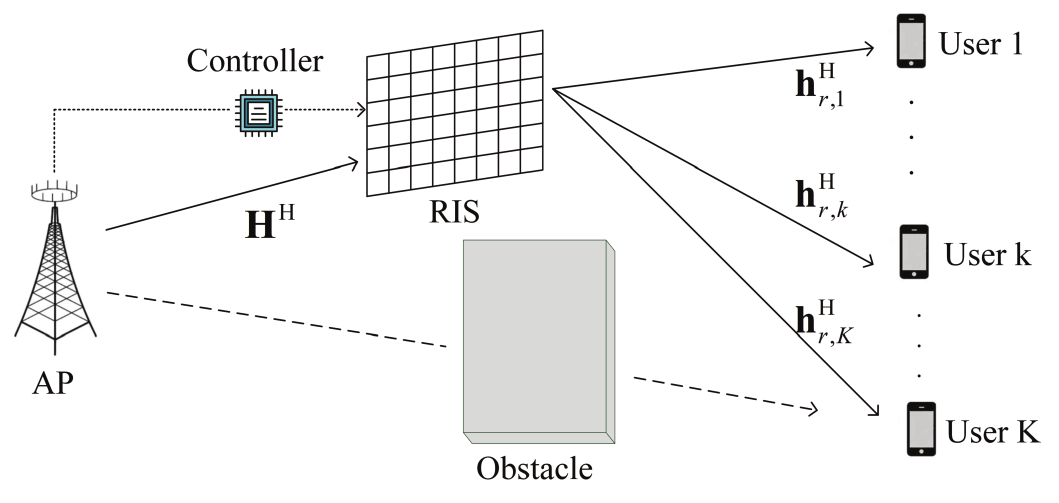}
\caption{An RIS-aided downlink communication system.}
\label{tu1}
\end{figure}

\subsection{System Model}

{\quad As depicted in Fig.~\ref{tu1}, we consider a RIS aided downlink MISO system, where an $M$-antenna BS serves $K$ single-users, whose indices are collected in $\mathcal{K}\buildrel \Delta \over = \left\{1,2, \cdots, K\right\}$. Like works~\cite{9066923,mu2021simultaneously,wang2021joint}, we assume that the direct communication link between the BS and users are blocked by obstacles. This scenario is a classic application for using the RIS to improve the coverage for users in the blind spots~\cite{wu2019towards}.} The AP adopts uniform linear array (ULA) with $M$ antennas spaced with half wave length, while the RIS contains $N$ reflection elements. 

Consider linear transmit beamforming at AP. Let digital beamforming matrix be $\mathbf{W}=\left[\mathbf{w}_{1}, \mathbf{w}_{2}, \cdots, \mathbf{w}_{K}\right] \in \mathbb{C}^{M \times K}$, where $\mathbf{w}_{k}$ denotes transmitting beamformer for user $k$.  The complex baseband transmit signal can be written as \vspace{-0.3cm}
\begin{equation}
\mathbf{x} = \mathbf{W}\mathbf{s},\vspace{-0.3cm}
\end{equation}
where  $\mathbf{s}=\left[s_{1}, s_{2}, \cdots, s_{K}\right]^{\mathrm{T}}$ stands for the transmitted data with $s_{k}$ representing the information symbol for user $k$ and satisfies $\mathbb{E}\left\{\mathbf{s}\mathbf{s}^{\mathrm{H}}\right\}= \mathbf{I}_{K}$. Besides, $\mathbf{W}$ satisfies transmit power constraint $\|\mathbf{W}\|_{\mathrm{F}}^{2} \le P$. $P$ is the total transmit power.

Signal received at user $k$ through RIS-aided channel is expressed as \vspace{-0.2cm}
\begin{equation}
\begin{small}
\begin{aligned}
y_{k} &= \mathbf{h}_{r, k}^{\mathrm{H}} \mathbf{\Phi} \mathbf{H}^{\mathrm{H}} \mathbf{W} \mathbf{s}+n_{k} \\
&=\underbrace{ \mathbf{h}_{r, k}^{\mathrm{H}} \mathbf{\Phi} \mathbf{H}^{\mathrm{H}} \mathbf{w}_{k} s_{k}}_{\text {useful signal }}+\underbrace{\sum_{j\ne k}\mathbf{h}_{r, k}^{\mathrm{H}} \mathbf{\Phi} \mathbf{H}^{\mathrm{H}} \mathbf{w}_{j} s_{j}+n_{k}}_{\text {interference and noise }}, \vspace{-0.3cm}
\end{aligned}
\end{small}
\end{equation}
where $\mathbf{H} \in \mathbb{C}^{M \times N}$ and $\mathbf{h}_{r, k} \in \mathbb{C}^{N \times 1}$ denote baseband equivalent channels from the RIS to AP, from the user $k$ to RIS, respectively. $n_{k} \sim \mathcal{C N}\left(0, \sigma_{0}^{2}\right)$ represents the additive white Gaussian noise (AWGN). $\mathbf{\Phi}=\operatorname{diag}(\bm{\theta }^{\mathrm{H}})$ is the passive beamforming matrix of the RIS, where $\bm{\theta}=\left[\exp \left(j \theta_{1}\right), \exp \left(j \theta_{2}\right), \ldots, \exp \left(j \theta_{N}\right)\right]^{\mathrm{H}}, \theta_{n} \in[0,2 \pi)$ represents phase shift introduced by RIS.

For simplicity, we ignore the non-linear coupling between reflection amplitude and phase shifts \cite{ref23} and take the common assumption that no energy loss occurs during RIS reflection. Hence, reflection amplitude of each RIS element is set to be $1$. In addition, signals reflected by RIS more than one time are neglected due to noticeable large-scale path loss \cite{ref6}.

Under the block fading channel assumption, the channel coefficients remain constant in the coherence block, but change independently from block to block. The ESE of user $k$ is given by \vspace{-0.1cm}
\begin{equation}\label{ESE}
\begin{small}
\begin{aligned}
&C_{k}(\mathbf{W}, \mathbf{\Phi}) =\mathbb{E}\left\{\log _{2}\left(1+\mathrm{SINR}_{k}\right)\right\} 
\\ &=\mathbb{E}\left\{\log _{2}\left(1+\frac{\left|\mathbf{h}_{r, k}^{\mathrm{H}} \mathbf{\Phi} \mathbf{H}^{\mathrm{H}} \mathbf{w}_{k}\right|^{2}}{\sum_{j \in\{\mathcal{K}\} / k} \left|\mathbf{h}_{r, k}^{\mathrm{H}} \mathbf{\Phi} \mathbf{H}^{\mathrm{H}} \mathbf{w}_{j}\right|^{2}+\sigma_{0}^{2}}\right)\right\},
\end{aligned}
\end{small}
\end{equation}
where multiuser interference is treated as noise.
\subsection{Channel Model}
\begin{figure}[htbp]
\centering
\includegraphics[width=3in]{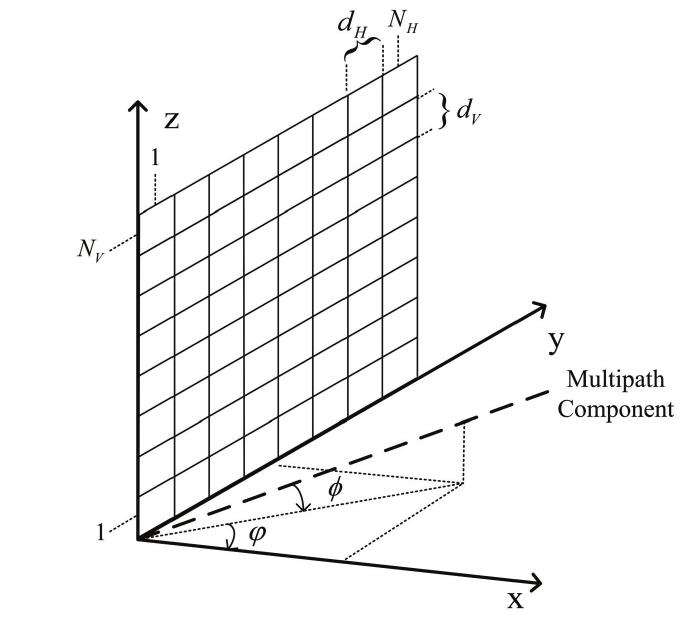}
\caption{The 3D geometry of a rectangular RIS.}
\label{tu2}
\end{figure}
The RIS is a rectangular surface consisting of $N_{\mathrm{H}}$ elements per row and $N_{\mathrm{V}}$ elements per column$(N=N_{\mathrm{H}}\times N_{\mathrm{V}})$. The setup is illustrated in Fig.~\ref{tu2} in a three-dimensional (3D) space with $\varphi \in \left[-\pi/2,\pi/2\right]$ and $\phi \in \left[-\pi/2,\pi/2\right]$ being the azimuth angle and the elevation angle, respectively. The reflection elements are indexed row-by-row and the location of the $n$th element, $n\in\mathcal{N}=\{1,2,\cdots,N\}$, is given by \vspace{-0.2cm}
\begin{equation}
\begin{small}
\mathbf{u}_{n}=\left[0, i_{\mathrm{H}}\left(n\right)d_{\mathrm{H}}, i_{\mathrm{V}}\left(n\right)d_{\mathrm{V}}\right]^{\mathrm{T}},\vspace{-0.2cm}
\end{small}
\end{equation}
where  $d_{\mathrm{H}}$ and $d_{\mathrm{V}}$ are the horizontal distance and the vertical distance between neighboring reflection elements of the RIS. $i_{\mathrm{H}}\left(n\right) = \mathrm{mod}\left(n-1,N_{\mathrm{H}}\right)$ and $i_{\mathrm{V}}\left(n\right) = \lfloor \left(n-1\right)/N_{\mathrm{H}}\rfloor$ are the column index and row index of the $n$th element, respectively. The RIS response vector corresponding to azimuth angle $\varphi$ and elevation angle $\phi$ is given by~\cite{ref20} \vspace{-0.2cm}
\begin{equation}
\mathbf{e}_{\text{RIS}}\left(\varphi,\phi\right)=\dfrac{1}{\sqrt{N}}\left[e^{j\mathbf{k}\left(\varphi,\phi\right)^{\mathrm{T}}\mathbf{u}_{1}}, \cdots, e^{j\mathbf{k}\left(\varphi,\phi\right)^{\mathrm{T}}\mathbf{u}_N}\right]^{\mathrm{T}} \vspace{-0.2cm}
\end{equation}

\noindent where $\mathbf{k}\left(\varphi,\phi\right) \in \mathbb{R}^{3\times1} $ is the wave vector \vspace{-0.2cm}
\begin{equation}
\mathbf{k}\left(\varphi,\phi\right)=\dfrac{2\pi}{\lambda}\left[\cos\left(\varphi\right)\sin\left(\phi\right), \cos\left(\phi\right)\sin\left(\varphi\right), \sin\left(\phi\right)\right]^{\mathrm{T}}.
\end{equation}

The response vector of the ULA at the AP corresponding to angle $\gamma \in \left[0,\pi\right]$ is given by~\cite{ref24} \vspace{-0.2cm}
\begin{equation}
\mathbf{e}_{\text{AP}}\left(\gamma\right)=\dfrac{1}{\sqrt{M}}\left[1,e^{j\pi\cos\left(\gamma\right)}, \cdots, e^{j\pi\left(M-1\right)\cos\left(\gamma\right)}\right]^{\mathrm{T}}.
\end{equation}

In practice, AP and RIS are usually deployed above surrounding objects. This means AP-RIS channels turn to be sparse for limited scattering objects and can be modeled as geometry channels with several dominant propagation paths~\cite{ref25}. The channel between them can be modeled as \vspace{-0.2cm}
\begin{equation}\label{H}
\begin{aligned}
\mathbf{H}=\sqrt{\frac{\beta_{H}M N}{\rho L}} \sum_{l=1}^{L} \rho_{l} \mathbf{e}_\text{AP}\left(\gamma_{l}\right) \mathbf{e}_\text{RIS}^{\mathrm{H}}\left(\varphi_{l}, \phi_{l}\right),
\end{aligned}
\vspace{-0.3cm}
\end{equation}

\noindent where $\gamma_{l}$, $\varphi_{l}$ and $\phi_{l}$ represent angle of arrival associated with the AP, azimuth angle of departure and elevation angle of departure associated with the RIS,  respectively. $L$ is the number of dominant paths, $\rho_{l}$ denotes the gain of the $l$th path and follows complex Gaussian distribution $\mathcal{C} \mathcal{N}\left(0, \sigma_{l}^{2}\right)$. $\rho=\sum_{l=1}^{L} \sigma_{l}^{2}$ is the normalization factor, $\beta_{H}$ stands for large scale path loss. Channel gains, angles of arrival and departure in \eqref{H} can be estimated based on existing channel estimation methods~\cite{ref26}. Since the positions of the RIS and the AP are fixed, the AP-RIS channel can be considered as quasi-static and is treated as a constant in our work~\cite{ref27}.

{Considering correlation between channel elements, each channel vector between RIS and users, \ie, $\mathbf{h}_{r,k}$, $\forall k\in\mathcal{K}$, is modeled as a realization of the CSCG distribution
\begin{equation}
\mathbf{h}_{r, k}\sim \mathcal{C N}\left(\mathbf{0}, \mathbf{R}_{r,k}\right),
\end{equation}
where $\mathbf{R}_{r, k}$ is the positive semi-deﬁnite covariance matrix which describes the spatial correlation between the channel components observed at the RIS from user $k$. The large-scale fading coefﬁcient between the RIS and user $k$ is determined by the normalized trace as \vspace{-0.2cm}
\begin{equation}\label{hrk}
\beta_{r,k}=\frac{1}{N}\tr\left(\mathbf{R}_{r,k}\right),\vspace{-0.2cm}
\end{equation}} 
\indent We assume there are limited scatterers in the environment which cause the signal of the signal reflected from the RIS to arrive at a given receiver via limited paths. This leads to a low-rank channel model which allows us to decompose the channel as\vspace{-0.2cm}
\begin{equation}\label{hrkd}
\mathbf{h}_{r,k}= \sum_{y=1}^{Y_k} \sqrt{\lambda_{k,y}} \tilde{h}_{k,y} [\mathbf{D}_k]_{y},\vspace{-0.2cm}
\end{equation}
where $Y_k$ is the rank of $\mathbf{R}_{r,k}$. $\mathbf{D}_k$ is the unitary matrix derived from the spectral decomposition (SD) of $\mathbf{R}_{r,k}$, i.e., $\mathbf{R}_{r,k}=\mathbf{D}_k\mathbf{\Lambda}_k\mathbf{D}_k^\mathrm{H}$, $\lambda_{k,1},\lambda_{k,2},\cdots,\lambda_{k,Y_k}$ eigenvalues of $\mathbf{R}_{r,k}$. The bi-orthogonal expansion renders expansion basis $[\mathbf{D}_k]_{1},[\mathbf{D}_k]_{2},\ldots, [\mathbf{D}_k]_{Y_k}$ being mutually orthogonal and expansion coefficients $\tilde{h}_{k,1},\tilde{h}_{k,2},\ldots,\tilde{h}_{k,Y_k}$ being mutually uncorrelated, i.e., $\tilde{\mathbf{h}}_k=[\tilde{h}_{k,1},\tilde{h}_{k,2},\ldots,\tilde{h}_{k,Y_k}]^\mathrm{T}\sim\mathcal{C N}\left(\mathbf{0}, \mathbf{I}\right)$.

\subsection{Problem Formulation}
\quad In this paper, our objective is to maximize the sum ESE of all users by designing the beamforming matrix  $\mathbf{W}$ at the AP and passive beamforming matrix $\bf{\Phi}$  at the RIS. The sum ESE maximization problem is formulated as \vspace{-0.4cm}
\begin{equation}
\label{P1}
\begin{aligned}
(\mathrm{P} 1): \max _{\mathbf{W}, \mathbf{\Phi}} & \sum_{k=1}^{K} C_{k}(\mathbf{W}, \mathbf{\Phi}) \\
\text { s.t. } &\|\mathbf{W}\|_{F}^{2}\le P \\
& \mathbf{\Phi}=\operatorname{diag}(\bm{\theta}^\mathrm{H}) \\
& 0 \leq \theta_{n}<2 \pi, n\in\mathcal{N}.
\end{aligned}
\vspace{-0.3cm}
\end{equation}
{\indent Due to the absence of closed-form analytical expression for the objective function of problem $(\mathrm{P} 1)$, it is quite difficult to carry out the joint beamforming design based on the ergodic spectrum efficiency.} Inspired by~\cite{ref11}, equationeq~\eqref{ESE} can be approximated as 
\begin{equation}\label{approximation}
\begin{aligned}
C_{k}&(\mathbf{W},\mathbf{\Phi})\approx \tilde C_k\left(\mathbf{W}, {\bf{\Phi }} \right)=\\
&\log _{2}\left( 1+\frac{  \mathbf{w}_j^{\mathrm{H}} \mathbf{H} \mathbf{\Phi}^{\mathrm{H}} \mathbf{R}_{r, k} \mathbf{\Phi} \mathbf{H}^{\mathrm{H}}\mathbf{w}_k}{\sum_{j \in\{\mathcal{K}\} / k}  \mathbf{w}_j^{\mathrm{H}} \mathbf{H} \mathbf{\Phi}^{\mathrm{H}} \mathbf{R}_{r, k} \mathbf{\Phi} \mathbf{H}^{\mathrm{H}}\mathbf{w}_j+\sigma_{0}^{2}}\right).
\end{aligned}
\end{equation}
{Problem $(\mathrm{P} 1)$ with the ergodic spectrum efficiency approximation is nonconvex due to the non-convex objective function and the non-convex RIS coefficients constraint.} In the following section, we try to solve this optimization problem based on ergodic spectrum efficiency approximation. 



\section{ Active Beamforming}

\quad In this section, we present two active beamforming algorithms, namely the SVD based active beamforming algorithm and the FP based beamforming algorithm. For the SVD based active beamforming algorithm, no alternative optimization is required for the joint beamforming, however this active beamforming strategy can lead to performance loss since the power is allocated evenly among all beamforming direction. For the FP based active beamforming algorithm, the active beamforming vectors are designed more carefully, however alternative optimization is required for joint beamforming. \vspace{-0.5cm}

\subsection{SVD based Active Beamforming Algorithm}
\quad  In this subsection, we present the SVD based active beamforming strategy. Since $\mathbf{W}$ and $\mathbf{\Phi}$ in problem (P1) are coupled with each other, existing works often adopt AO method to design $\mathbf{W}$ and $\mathbf{\Phi}$. However, AO algorithm needs to repeatedly solve complex optimization problems, thus brings about large time and energy consumption. It is appealing to decouple $\mathbf{W}$ and $\mathbf{\Phi}$ to obtain suboptimal algorithm with low complexity.\\
\noindent Under the circumstance of obstructed AP-user links, it is intuitively true that the AP sends signals to RIS for assisting communication. Note that AP-RIS-user link is a two-hop channel: in the first hop, directions of beamforming are determined by the AP-RIS channel, whereas the AP power allocation strategy is subject to the RIS  phase \cite{ref21}; in the second hop, the RIS phase is in turn affected by the AP beamforming matrix. Since path gains are independent and identically distributed (i.i.d.)  random variables in AP-RIS channel, it is reasonable to allocate the power evenly among beamforming directions. Under this circumstance, the AP beamforming matrix is determined by the AP-RIS channel. Hence AP beamforming matrix $\mathbf{W}$ and RIS phase $\mathbf{\Phi}$ are decoupled and we can confront $\mathbf{W}$ and $\mathbf{\Phi}$ optimization problems one by one.\\
\indent Note that optimizing AP beamforming matrix $\mathbf{W}$ for AP-RIS link is a classic point to point MIMO beamforming problem, which can be solved by conducting SVD of AP-RIS channel $\mathbf{H}$ and set the direction of beamforming vectors to be the columns of the left unitary matrix derived from the SVD \cite{ref28}.\\ 
\indent The SVD of AP-RIS channel $\mathbf{H}$ is  \vspace{-0.3cm}
\begin{equation}
{\bf{H = U\Sigma }}{{\bf{V}}^{\rm{H}}}, \vspace{-0.3cm}
\end{equation}
where $\mathbf{V}$ and $\mathbf{U}$ are the unitary matrices derived from SVD. $\mathbf{\Sigma}$ is a rectangular diagonal matrix, i.e.,

\begin{equation} 
\mathbf{\Sigma}=\left[ 
\begin{array}{cccccc} 
\delta _1 & 0 & \cdots & 0 & \cdots & 0\\ 
0 & \delta _2 & \cdots & 0 & \cdots & 0\\ 
0 & 0 & \ddots & 0 & \cdots & 0\\ 
0 & 0 & \cdots & \delta _M & \cdots & 0\\ 
\end{array}
\right] 
\end{equation}
with ${\delta _1},{\delta _2}, \ldots ,{\delta _M}$ denote the descending ordered singular values of matrix $\mathbf{H}$. Here we omit the path loss $\beta_{H}$ for brevity.

Based on aforementioned power allocation strategy, we set the beamforming vectors to be the columns of the left unitary matrix $\mathbf{U}$, the AP beamforming matrix is given by \vspace{-0.4cm}
\begin{equation}\label{SVD}
\mathbf{W}=\sqrt{P}\Big[[\mathbf{U}]_{1},[\mathbf{U}]_{2},\cdots,[\mathbf{U}]_{K}\Big]\vspace{-0.4cm}
\end{equation}
where $[\mathbf{U}]_{k}$  denotes the $k$th column for matrix $\mathbf{U}$ and serves as the beamforming vector for the  $k$th user. We design $\mathbf{W}$ only according to $\mathbf{H}$ to avoid using alternative optimization methods. It is not the optimal option, but it functions well to transmit information from the AP to the RIS.

\subsection{FP based Active Beamforming Algorithm}
\quad  In this subsection, we present the FP based active beamforming strategy.

First, for given passive beamforming matrix $\bm{\Phi}$, problem (P1) can be reformulated as
\begin{equation}
\label{P2}
\begin{aligned}
(\mathrm{P} 2): \max _{\mathbf{W}} & \sum_{k=1}^{K} \mathrm{log}\left(1+\frac{\mathbf{w}_k^\mathrm{H}\mathbf{A}_k\mathbf{w}_k}{\sum_{j \in\{\mathcal{K}\} / k}\mathbf{w}_j^\mathrm{H}\mathbf{A}_k\mathbf{w}_j+\sigma_{0}^{2}}\right)\\
\text { s.t. } &\|\mathbf{W}\|_{F}^{2}\le P.
\end{aligned}
\end{equation}
where $\mathbf{A}_k=\mathbf{H} \mathbf{\Phi}^{\mathrm{H}} \mathbf{R}_{r, k} \mathbf{\Phi} \mathbf{H}^{\mathrm{H}}$.

By using the lagrangian dual transform \cite{guo2020weighted} and introducing auxiliary variables $\gamma_k, k=1,2,\cdots,K$, (P2) can be rewritten as
\begin{equation}
\label{P3}
\begin{aligned}
(\mathrm{P} 3): \max _{\mathbf{W}, \bm{\gamma}} & \sum_{k=1}^{K} f_k\left(\mathbf{W}, \bm{\gamma}\right)\\
\text { s.t. } &\|\mathbf{W}\|_{F}^{2}\le P.
\end{aligned}
\end{equation}
where $ f_k\left(\mathbf{W}, \bm{\gamma}\right)=\frac{\left(1+\gamma_k\right)\mathbf{w}_k^\mathrm{H}\mathbf{A}_k\mathbf{w}_k}{\sum_j\mathbf{w}_j^\mathrm{H}\mathbf{A}_k\mathbf{w}_j+\sigma_{0}^{2}}+\mathrm{log}\left(1+\gamma_k\right)-\gamma_k$ is the lagrangian dual function. 

The two problems are equivalent in the sense that $\mathbf{W}$ is the solution to problem (P2) if and only if it is the solution to problem (P3), and the optimal objective values of these two problems are also equal.

Substituting equation \eqref{hrk} and \eqref{hrkd} into $ f_k\left(\mathbf{W}, \bm{\gamma}\right)$, we divide the first item of $ f_k\left(\mathbf{W}, \bm{\gamma}\right)$ into $Y$ parts, i.e.,
\begin{equation}
\begin{aligned}
&\frac{\left(1+\gamma_k\right)\mathbf{w}_k^\mathrm{H}\mathbf{A}_k\mathbf{w}_k}{\sum_j\mathbf{w}_j^\mathrm{H}\mathbf{A}_k\mathbf{w}_j+\sigma_{0}^{2}}=\\
&\sum_y^Y\frac{\lambda_k\left(1+\gamma_k\right)\mathbf{w}_k^\mathrm{H}[\mathbf{D}_k]_{y}[\mathbf{D}_k]_{y}^\mathrm{H}\mathbf{w}_k}{\sum_j\mathbf{w}_j^\mathrm{H}\mathbf{A}_k\mathbf{w}_j+\sigma_{0}^{2}},
\end{aligned}
\end{equation}
where, without loss of generality, we set $Y_1=Y_2=\cdots=Y_K=Y$. 

The optimal $\gamma_k$ can be obtained by setting $\partial {f_k}/\partial {\gamma _k}$ to zero. With calculated optimal $\bm{\gamma}$, the problem (P3) becomes a multi-ratio FP problem which can be expressed as
\begin{equation}
\label{P4}
\begin{aligned}
(\mathrm{P} 4): \max _{\mathbf{W}} &  f\left(\mathbf{W}\right)\\
\text { s.t. } &\|\mathbf{W}\|_{F}^{2}\le P.
\end{aligned}
\end{equation}
where $ f\left(\mathbf{W}\right)=\sum_k\sum_y\frac{\lambda_k\left(1+\gamma_k\right)\mathbf{w}_k^\mathrm{H}[\mathbf{D}_k]_{y}[\mathbf{D}_k]_{y}^\mathrm{H}\mathbf{w}_k}{\sum_j\mathbf{w}_j^\mathrm{H}\mathbf{A}_k\mathbf{w}_j+\sigma_{0}^{2}}$. 

Using the quadratic transform proposed in \cite{shen2018fractional}, the stationary solution of $\mathbf{W}$ can be obtained with the following
iterative updating rule
\begin{equation}\label{qky}
\begin{aligned}
q_{ky}=\frac{\sqrt{\lambda_k\left(1+\gamma_k\right)}\mathbf{w}_k^\mathrm{H}[\mathbf{D}_k]_{y}}{\sum_j\mathbf{w}_j^\mathrm{H}\mathbf{A}_k\mathbf{w}_j+\sigma_{0}^{2}},
\end{aligned}
\end{equation}
\begin{equation}\label{wk}
\begin{aligned}
\mathbf{w}_k=&\sqrt{\left(1+\gamma_k\right)}\left(\lambda \mathbf{I}+\sum_k\sum_y|q_{ky}|^2\mathbf{A}_k\right)^{-1}\\
&\times \left(\sum_yq_{ky}^{*}[\mathbf{D}_k]_y\right),
\end{aligned}
\end{equation}
where $\mathbf{Q}\in\mathbb{C}^{K\times Y}$ is an auxiliary matrix, $q_{ky}$ is the element on the $k$-th row and $y$-th column of $\mathbf{Q}$, $\lambda$ is the dual variable introduced for the transimit power constraint, which can be determined efficiently through bisection search.
\begin{algorithm}[!htbp]
\caption{The FP based Active Beamforming Algorithm with Given Passive Beamforming Matrix $\mathbf{\Phi}$}
\begin{algorithmic}[1]
\STATE Initialize $\mathbf{W}^0$ and set the iteration index $n=0$
\REPEAT 
\STATE $n=n+1$
\FOR{$ k=1:K $}               
\STATE Calculate $\gamma_k^n$ by setting $\partial {f_k}/\partial {\gamma _k}$ to zero
\ENDFOR
\FOR{$ k=1:K $}               
\FOR{$ y=1:Y $}               
\STATE Calculate $q_{ky}^n$ according to \eqref{qky}
\ENDFOR
\ENDFOR
\STATE Calculate $\mathbf{w}_{k}^n$ according to \eqref{wk}
\UNTIL {the objective function of (P2) converges or $n$ reaches the maximum
iteration number}
\end{algorithmic}
\label{algorithm1}
\end{algorithm}
The FP based active beamforming algorithm with given passive beamforming matrix $\mathbf{\Phi}$ is summarized in Algorithm~\ref{algorithm1}. For the n-th iteration, in Steps $3$-$5$, the optimal auxiliary variable vector $\bm{\gamma}^n$ for the lagrangian dual transform is calculated. In Steps $7$-$11$, the optimal auxiliary matrix $\mathbf{Q}^n$ for the quadratic transform is calculated. In Steps $12$, the active beamforming matrix $\mathbf{W}^n$ for the n-th iteration is calculated. The iteration will repeat until the objective function of (P2) converges or n reaches the maximum iteration number.
\section{PASSIVE BEAMFORMING}
\quad In this section, the multi-user passive beamforming problem for given $\mathbf{W}$ is investigated. \\
\indent When the active beamforming matrix is given using either algorithm from section III, the sum ESE maximization problem (P1) is reformulated as \vspace{-0.2cm}
\begin{equation}
\label{P5}
\begin{aligned}
(\mathrm{P} 5): &\max _{{\bm{\theta }}} \sum_{k=1}^{K}\tilde f_{k}({\bm{\theta }}) \\
& \begin{aligned}
&0 \leq \theta_{n}<2 \pi,  n\in\mathcal{N}, \vspace{-0.2cm}
\end{aligned}
\end{aligned}
\end{equation}

\noindent where \vspace{-0.2cm}
\begin{equation}
\begin{aligned}
\tilde f_{k}\left( {\bm{\theta }} \right)=\log _{2}\left( 1+\frac{\bm{\theta} ^{\mathrm{H}}\mathbf{R}_k^k \bm{\theta}}{ \bm{\theta} ^{\mathrm{H}}\left(\sum_{j \in\{\mathcal{K}\} / k}\mathbf{R}_k^j+\sigma_{0}^{2}\mathbf{I}\right) \bm{\theta}}\right)
\end{aligned}
\end{equation}
with $\mathbf{R}_k^j=  \mathrm{diag}\left(\mathbf{H}^{\mathrm{H}}\mathbf{w}_{j} \right)^{\mathrm{H}}\mathbf{R}_{r,k} \mathrm{diag}\left(\mathbf{H}^{\mathrm{H}}\mathbf{w}_{j} \right)$.

Problem (P5) can be solved by using gradient descent until converging to a stationary point. The gradient of $\tilde f_{k}\left( {\bm{\theta }} \right)$ is given by \vspace{-0.2cm}
\begin{equation}
\begin{aligned}\label{GD}
&\frac{\partial {\tilde f_k\left(\bm{\theta}\right)}}{\partial \bm{\theta}^{*}}=\\
&\frac{\frac{\mathbf{R}_k^k \bm{\theta}}{\bm{\theta} ^{\mathrm{H}}\left(\sum_{j \in\{\mathcal{K}\} / k}\mathbf{R}_k^j+\sigma_{0}^{2}\mathbf{I}\right) \bm{\theta}}+\frac{\bm{\theta} ^{\mathrm{H}}\mathbf{R}_k^k \bm{\theta}\left(\sum_{j \in\{\mathcal{K}\} / k}\mathbf{R}_k^j+\sigma_{0}^{2}\mathbf{I}\right) \bm{\theta}}{\left(\bm{\theta} ^{\mathrm{H}}\left(\sum_{j \in\{\mathcal{K}\} / k}\mathbf{R}_k^j+\sigma_{0}^{2}\mathbf{I}\right) \bm{\theta}\right)^2}}{\mathrm{ln}\left(2\right)\left(1+\frac{\bm{\theta} ^{\mathrm{H}}\mathbf{R}_k^k \bm{\theta}}{\bm{\theta} ^{\mathrm{H}}\left(\sum_{j \in\{\mathcal{K}\} / k}\mathbf{R}_k^j+\sigma_{0}^{2}\mathbf{I}\right) \bm{\theta}}\right)}.
\end{aligned}
\end{equation}

Assume the variable in the $t$-th iteration is $\bm{\theta}^t$. Then, the variable $\bm{\theta}^{t+1}$ in the next iteration is given by \vspace{-0.2cm}
\begin{equation}\label{theta}
\begin{aligned}
\tilde{\bm{\theta}}^{t}=\bm{\theta}^t+\mu \sum_k^K \frac{\partial {\tilde f_k\left(\bm{\theta}\right)}}{\partial \bm{\theta}^{*}}|_{\bm{\theta}=\bm{\theta}^{t}}, 
\end{aligned}
\end{equation}
\vspace{-0.4cm}
\begin{equation}\label{thetaunit}
\begin{aligned}
{\bm{\theta}}^{t}=\mathrm{exp}\left(j\mathrm{arg}\left(\tilde{\bm{\theta}}^{t+1}\right)\right),
\end{aligned}
\end{equation}
where the step size $\mu$ is chosen by backtracking line search \cite{papazafeiropoulos2021asymptotic}. Equation \eqref{thetaunit} is a projection operation for meeting the unit modulus constraint.

\begin{algorithm}[!htbp]
\caption{The GD based passive Beamforming Algorithm with Given Active Beamforming Matrix $\mathbf{W}$}
\begin{algorithmic}[1]
\STATE Initialize $\bm{\theta}^0$ and set the iteration index $t=0$
\REPEAT 
\STATE $t=t+1$
\FOR{$ k=1:K $}               
\STATE Calculate $\frac{\partial {\tilde f_k\left(\bm{\theta}\right)}}{\partial \bm{\theta}^{*}}|_{\bm{\theta}=\bm{\theta}^{t-1}}$ according to \eqref{GD}
\ENDFOR  
\STATE Calculate $\tilde{\bm{\theta}}^{t+1}$ according to \eqref{theta}
\STATE Calculate ${\bm{\theta}}^{t+1}$ according to \eqref{thetaunit}
\UNTIL {the objective function of (P5) converges or $n$ reaches the maximum
iteration number}
\end{algorithmic}
\label{algorithm2}
\end{algorithm}
The GD based passive beamforming algorithm with given active beamforming matrix $\mathbf{W}$ is summarized in Algorithm~\ref{algorithm2}. For the t-th iteration, in Steps $3$-$5$, the gradient of $\sum_k^K\tilde f_{k}\left( {\bm{\theta }} \right)$ is calculated. In Steps $7$-$8$, the passive beamforming ${\bm{\theta}}^{t}$ vector for the t-th iteration is calculated. The iteration will repeat until the objective function of (P5) converges or $n$ reaches the maximum iteration number.
\section{Overall Algorithms and Pilot Overhead}
{\quad For the SVD-GD algorithm, the the active beamforming matrix $\mathbf{W}$ is calculated  with \eqref{SVD} and the passive beamforming matrix $\mathbf{\Phi}$ is given by solving problem $(\mathrm{P} 2)$ with Algorithm~\ref{algorithm1}. Thus no alternative optimization is required for the joint beamforming. However, the SVD based active beamforming can lead to performance loss since the power is allocated evenly among all beamforming direction.\\
\indent For the FP-GD algorithm, which is summarized in Algorithm~\ref{algorithm3}, the active beamforming vectors are designed more carefully. However, alternative optimization is required for joint beamforming. }
\begin{algorithm}[!htbp]
{\caption{{The Proposed FP-GD Algorithm}}\label{algorithm3}
\begin{algorithmic}[1]
\STATE {Initialize feasible $\mathbf{W}$ that satisfies the power constraint of problem $(\mathrm{P} 1)$.}
\STATE {{\bf{repeat:}}}
\STATE {\quad Given $\mathbf{W}$, update the passive beamforming matrix $\mathbf{\Phi}$ by solving problem $(\mathrm{P} 5)$ with Algorithm~\ref{algorithm2}.}
\vspace{-0.4cm}
\STATE {\quad Given $\mathbf{\Phi}$, update the active beamforming matrix $\mathbf{W}$ by solving problem $(\mathrm{P} 2)$ with Algorithm~\ref{algorithm1}.}
\STATE  {{\bf until} the fractional increase of the objective function
value of problem $(\mathrm{P} 1)$ is below a predefined threshold.}
\end{algorithmic}}
\end{algorithm}

\indent The quasi-static AP-RIS channel is estimated using methods in \cite{ref27}, with the minimum required pilot overhead $\tau_1=2\left(N+1\right)$. The approach to estimate the correlation matrices for RIS-user channels is to form the sample correlation matrix. For user $k$, suppose the AP has made independent observations of $\mathbf{h}_{r,k}$ in $S$ coherence blocks, each of which consists of pilot overhead $\tau$, where $\tau$ is the required signaling overhead of obtaining one realization of $\mathbf{h}_{r,k}$. The sample correlation matrix for user $k$ is\vspace{-0.4cm}
\begin{equation}
{{\bf{R}}_{r,k}^{sample}} = \frac{1}{S}\sum\limits_{i = 1}^{S}{{\bf{h}}_{r,k}^{i}\left({\bf{h}}_{r,k}^i\right)^{\rm{H}}}, \vspace{-0.4cm}
\end{equation}

\noindent where $\mathbf{h}_{r,k}^{i}$ denotes the observations of $\mathbf{h}_{r,k}$ in coherence blocks $i$.\\
\indent For each element of ${\bf{R}}_{r,k}^{sample}$, the law of large numbers implies that the sample variance converges (almost surely) to the true variance of corresponding element of ${\bf{R}}_{r,k}$. The standard deviation of the sample variance decays as ${1 \mathord{\left/
 {\vphantom {1 {\sqrt S }}} \right.
 \kern-\nulldelimiterspace} {\sqrt S }}$, thus a small number of observation blocks $S$ is sufficient to get a good variance estimate. The required signaling overhead of obtaining all the correlation matrix ${\bf{R}}_{r,k}, k\in \left\{1,2,\dots,K\right\}$ is $\tau_2=KS\tau$.

The pilot overhead is calculated by \vspace{-0.3cm}
\begin{equation}
\tau=\tau_1+\tau_2 = 2\left(N-1\right) + KS\tau. \vspace{-0.3cm}
\end{equation}
\indent Since the quasi-static RIS-BS channel and the slow-varying statistical CSI for RIS-user channels are estimated in a larger timescale than instantaneous CSI, the average pilot overhead associated with proposed algorithm can be reduced from a long-term perspective.
\section{SIMULATION RESULTS}
\label{SIMULATION-RESULTS}
\quad In this section, simulation results are provided to validate the effectiveness of proposed algorithms and draw useful insights.

\subsection{Simulation Setup}
\begin{figure}[!htbp]
\centering
\includegraphics[width=3.5in]{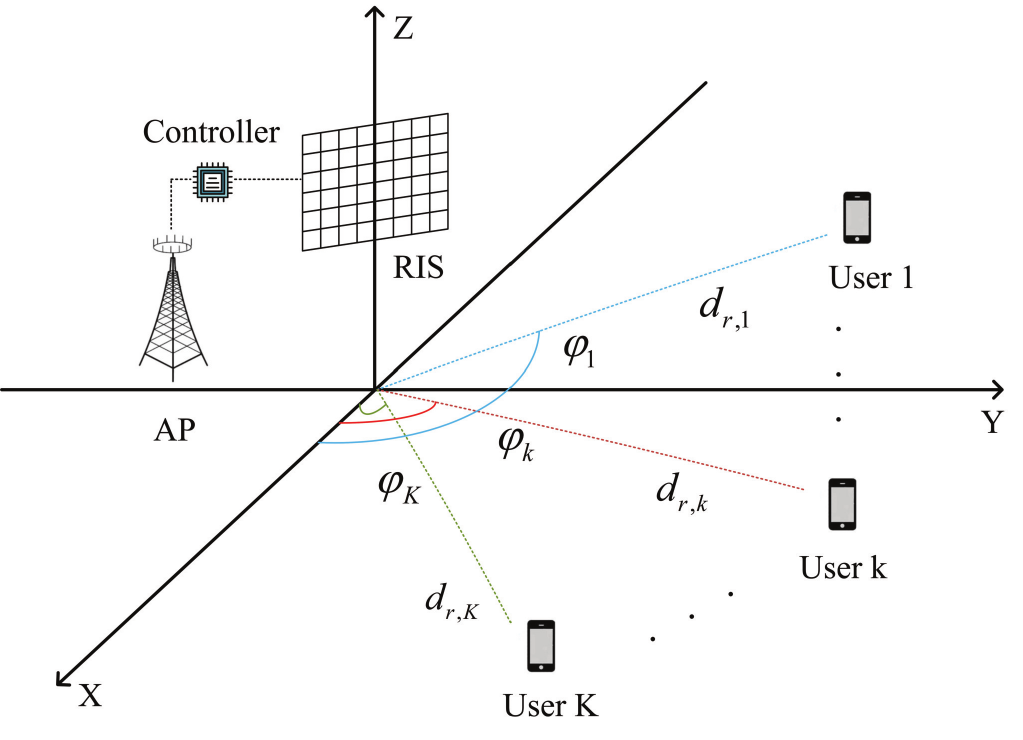}
\caption{Simulation setup.}\label{tu3}
\end{figure}

As shown in Fig.~\ref{tu3}, a 3D coordinate system is adopted where the locations of AP and RIS are set at (0 m, -15 m, 15 m) and (0 m, 0 m, 15 m), respectively. User $k$ is located in $x$-$y$ plane with distance $d_{r,k}$ and azimuth angle $\varphi_{k}$ relative to coordinate origin.

\begin{figure}[!htbp]
\centering
\includegraphics[width=3.5in]{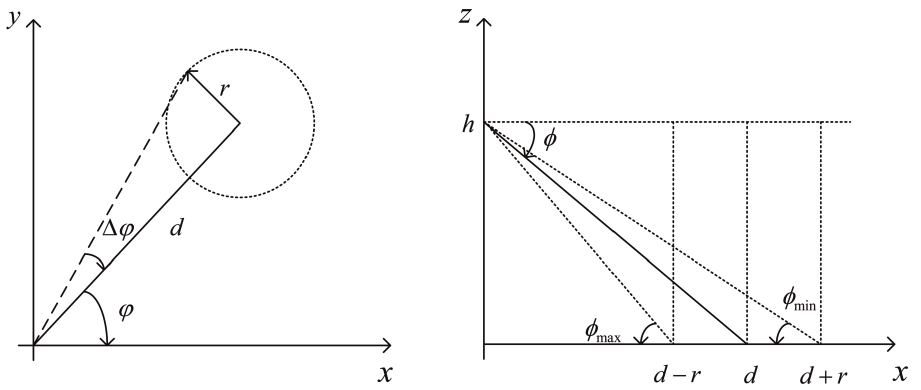}
\caption{The 3D one-ring model.}\label{tu4}
\end{figure}

As shown in Fig.~\ref{tu4}, consider the classic 3D one-ring model which assumes that scatters exist only on a ring of radius $r$ around user \cite{ref31,ref32}. The height of RIS is $h$. For a user locating in $x$-$y$ plane with azimuth distance $d$ and azimuth angle $\varphi$ relative to coordinate origin, azimuth angular spread of RIS-user channel is \vspace{-0.2cm}
\begin{equation}
\Delta \varphi=\arctan \left(\frac{r}{d}\right). \vspace{-0.2cm}
\end{equation}

Elevation angle and elevation angle spread of RIS-user channel are
\begin{equation}
\phi=\frac{1}{2}\left(\arctan \left(\frac{h}{d-r}\right)+\arctan \left(\frac{h}{d+r}\right)\right)
\end{equation}
and
\begin{equation}
\Delta\phi=\frac{1}{2}\left(\arctan \left(\frac{h}{d-r}\right)-\arctan \left(\frac{h}{d+r}\right)\right)
\end{equation}
respectively. Following \cite{ref20}, use formulation (7.17) in \cite{ref29} to calculate spatial correlation matrix based on aforementioned angular domain parameters. Particularly, the Laplacian distribution is used to characterize angular distribution for RIS-user channels to make results more practical \cite{ref33}.

Channel path-loss coefficient is modeled as

\begin{equation}
\beta(d_l)=C_{0}\left(\frac{d_l}{D_{0}}\right)^{-\alpha},
\end{equation}
where $C_{0}=-30\mathrm{dB}$ is the path loss at reference distance $D_{0}=1\mathrm{m}$. $d_l$ represents link distance and $\alpha$ denotes path loss exponent. The path loss exponents of AP-RIS link and RIS-user links are $\alpha_\mathrm{H}=2.2$ and $\alpha_\mathrm{r}=3$, respectively. Other system parameters are set as follows unless otherwise specified: AP antenna spacing is $\lambda / 2$, while RIS elements separation is $\lambda / 8 $ \cite{ref19},  $\sigma_{0}^{2}=-93\mathrm{dBm}$, $M = 16$, $N_\mathrm{H}= N_\mathrm{V}=20$ $K=3$. The unsymmetric user location setup is listed in Table~\ref{tabel1}, where the RIS-user channels are statistically equally strong. The symmetric user location setup is listed in Table~\ref{tabel2}, where the RIS-user channels are not statistically equally strong. All results are averaged over $10^{3}$ independent channel realizations.
\begin{table}[!htbp]
\begin{scriptsize}
\caption{The unsymmetric user location setup}\label{tabel1}
\centering
\begin{tabular}{|c||c||c||c|}
\hline
~ & azimuth angle & scatters ring radius & azimuth distance\\
\hline
User 1 & $\pi/3$ & 30m & 80m\\
\hline
User 2 & $\pi/4$ & 40m & 90m\\
\hline
User 3 & $\pi/6$ & 50m & 100m\\
\hline
\end{tabular}
\end{scriptsize}
\end{table}
\begin{table}[!htbp]
\begin{scriptsize}
\caption{The symmetric user location setup}\label{tabel2}
\centering
\begin{tabular}{|c||c||c||c|}
\hline
~ &azimuth angle&scatters ring radius&azimuth distance\\
\hline
User 1 & $\pi/3$ & 50m & 100m\\
\hline
User 2 & $\pi/4$ & 50m & 100m\\
\hline
User 3 & $\pi/6$ & 50m & 100m\\
\hline
\end{tabular}
\end{scriptsize}
\end{table}
\subsection{Simulation Results}
\begin{figure}[!htbp]
\centering
\includegraphics[width=3.5in]{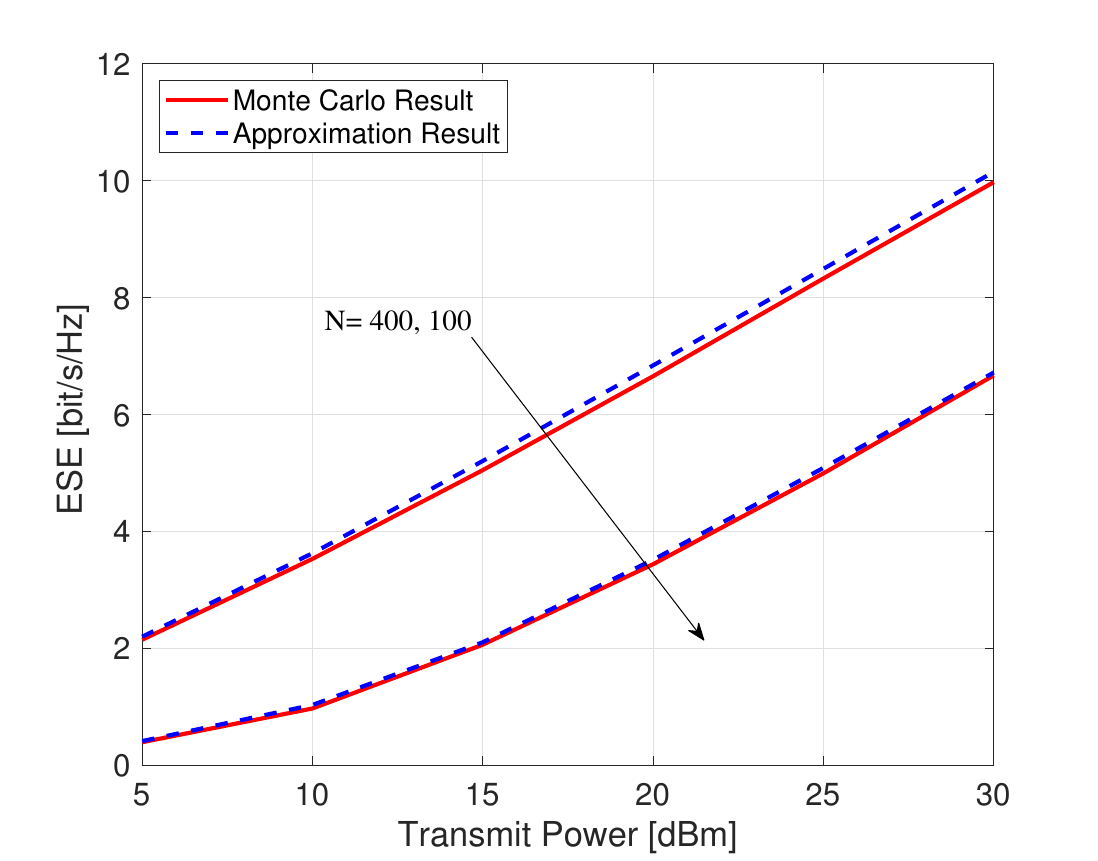}
\caption{The validation of the ESE approximation.}\label{tu5}
\end{figure}
\quad To verify the accuracy of the approximation in \eqref{approximation}, Fig.~\ref{tu5} compares the approximation result with the ESE, which is evaluated through the computationally expensive Monte Carlo method. As can be seen from Fig.~\ref{tu5}, the differences between the Monte Carlo results and the approximation results are almost negligible. The simulation results validate the accuracy of \eqref{approximation}.

\begin{figure}[!htbp]
\centering
\includegraphics[width=3.5in]{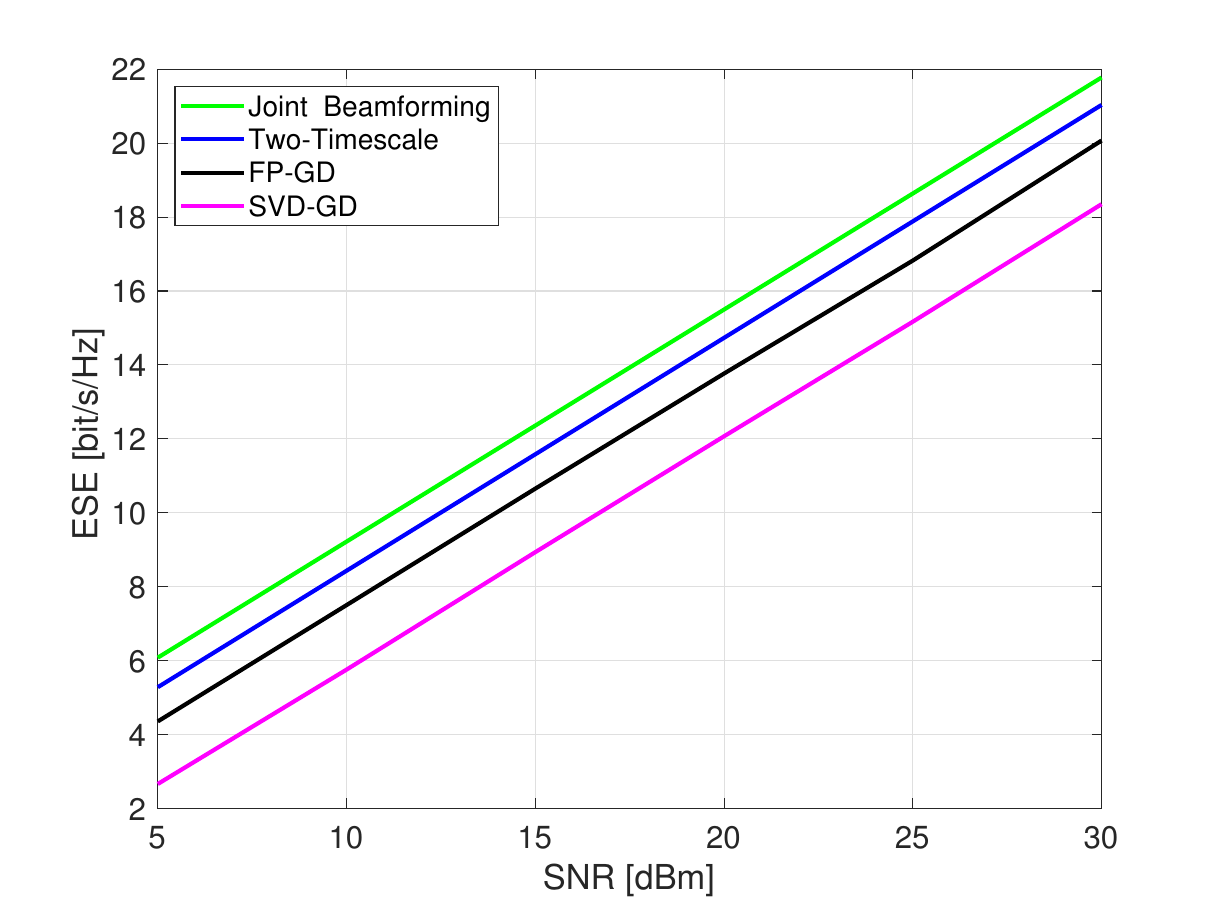}
\caption{AP transmit power versus ESE under the user setup in Table~\ref{tabel1}.}\label{tu6}
\end{figure}
We compare the performance of the FP-GD algorithm and the SVD-GD algorithm (the FP-GD algorithm takes the FP active beamforming method and the GD passive beamforming method, while the SVD-GD algorithm takes the SVD active beamforming method and the GD passive beamforming method) with following algorithms in Fig.~\ref{tu6} :

\begin{enumerate}
\item{Joint Beamforming Algorithm: $\mathbf{W}$ and $\mathbf{\Phi}$ are optimized using the joint beamforming scheme in \cite{ref6} based on I-CSI.}
\item{Two Time-scale Algorithm: $\mathbf{W}$ and $\mathbf{\Phi}$ are optimized using two time-scale beamforming scheme in \cite{ref14}.}
\end{enumerate}

Fig.~\ref{tu6} illustrates the sum ESE versus the AP transmit power for the user location setup in Table~\ref{tabel1}. The proposed S-CSI based algorithms are compared with the state-of-the-art I-CSI based joint beamforming scheme in~\cite{ref6} and two-timescale based beamforming scheme in~\cite{ref14}. The performance of the joint beamforming algorithm in~\cite{ref6} is the best since it assumes that perfect CSI is available at the AP. The performance of the two-timescale based algorithm in~\cite{ref14} also outperforms the proposed scheme because it acquires more CSI than the proposed scheme. The proposed schemes show a slightly degraded performance with reduced signaling burden since the proposed schemes only requires S-CSI. Besides, the FP-GD algorithm outperforms the SVD-GD algorithm since the transmit power for each user is designed more carefully in the FP-GD algorithm with alternative optimization.

\begin{figure}[h]
\centering
\includegraphics[width=3.5in]{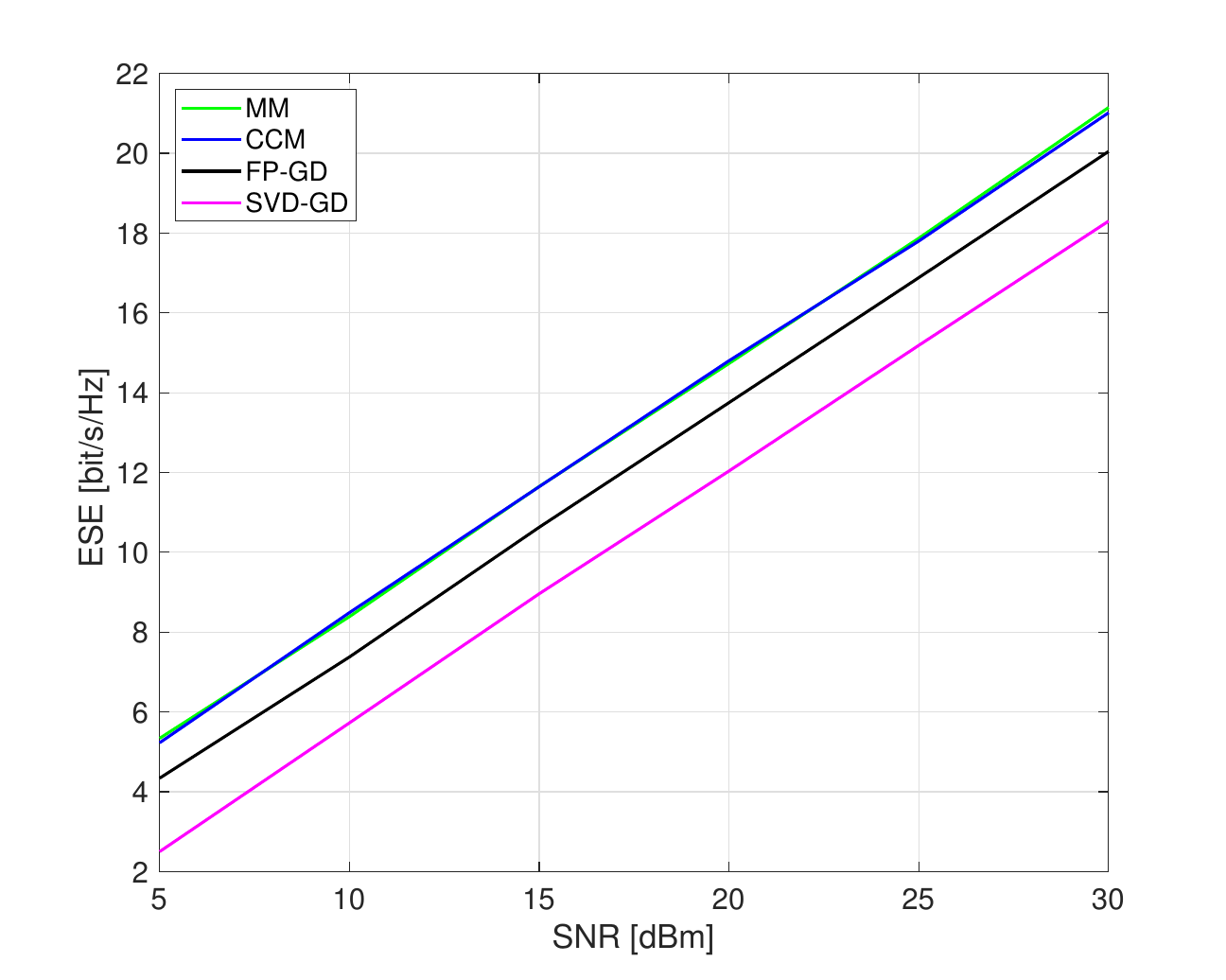}
\caption{AP transmit power versus ESE under the user setup in Table~\ref{tabel1}.}\label{tu7}
\end{figure}

For passive beamforming, we compare the performance of GD algorithm with following algorithms in Fig.~\ref{tu7} :
\begin{enumerate}
\item{Majorization-Minimization (MM) Algorithm: $\mathbf{W}$ and $\mathbf{\Phi}$ are optimized using the joint beamforming scheme in \cite{pan2020multicell}.}
\item{Complex Circle Manifold (CCM) Algorithm: $\mathbf{W}$ and $\mathbf{\Phi}$ are optimized using the joint beamforming scheme in \cite{yu2019miso}.}
\end{enumerate}

Fig.~\ref{tu7} illustrates the sum ESE versus the AP transmit power for the user location setup in Table 1. The proposed GD algorithm is compared with the the MM algorithm in \cite{pan2020multicell} and the CCM algorithm in \cite{yu2019miso}. The performance of two benchmark algorithms is very similar. Although the benchmark algorithms outperform the proposed GD algorithm. The complexity of the MM algorithm and the CCM algorithm is $\mathcal{O}(N^3)$, while the complexity of the GD algorithm is only $\mathcal{O}(N^2)$. (In RIS-aided system, the number of RIS elements ($N$) is usually much greater than the number of users ($K$) and the number of BS antennas ($M$). Thus, the computational complexity of the proposed algorithm is dominated by $N$). The proposed schemes show a slightly degraded performance with reduced computational complexity. Besides, the FP-GD algorithm outperforms the SVD-GD algorithm, which is inline with previous discussion.

\begin{figure}[!htbp]
\centering
\includegraphics[width=3.5in]{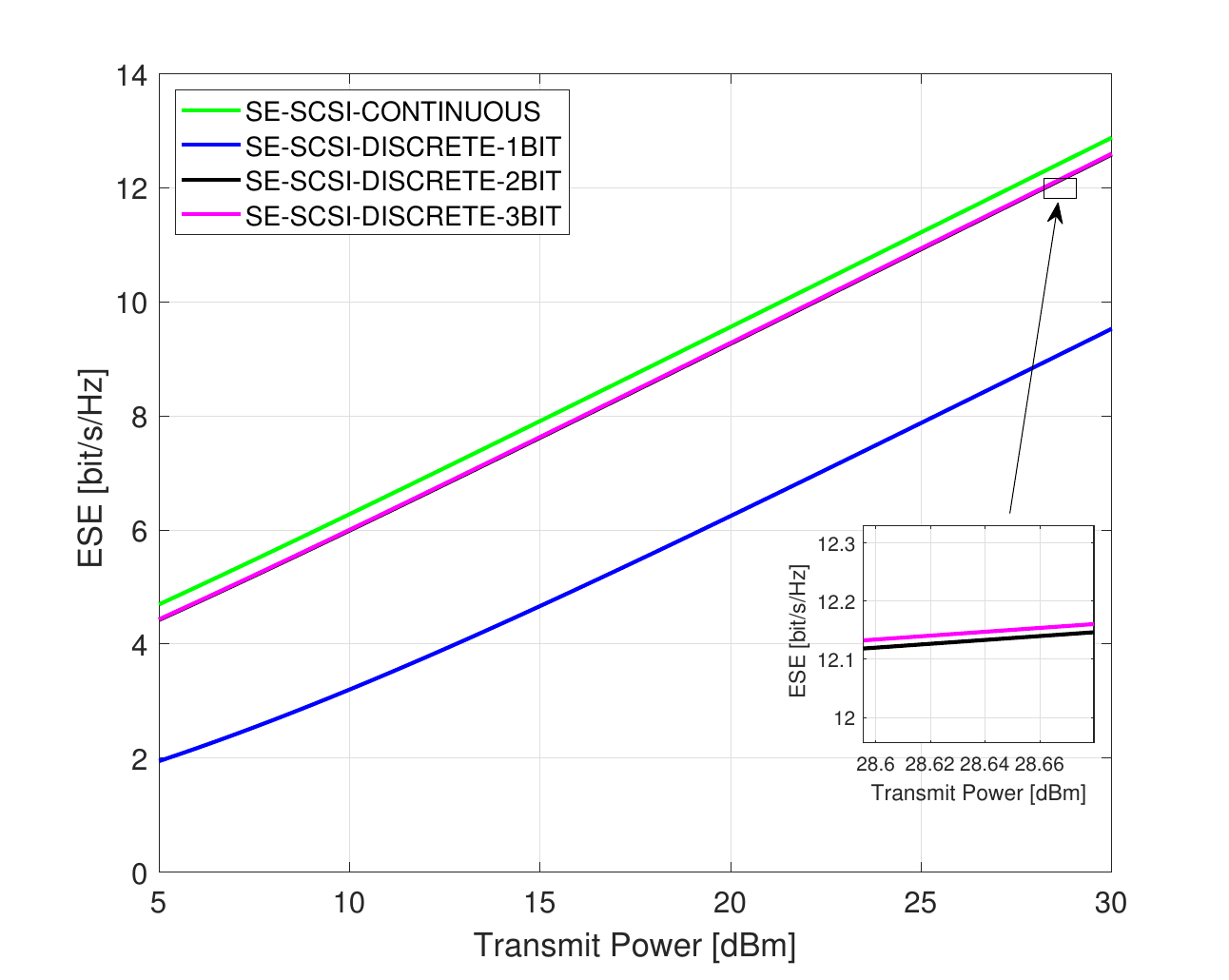}
\caption{AP transmit power versus ESE for different quantification.}\label{tu8}
\end{figure}

Fig.~\ref{tu8} illustrates user ESE versus AP transmit power for various quantization under the user setup in Table 1. Phase shift of each RIS reflection element is quantized by a $b$-bits quantizer, and the set of discrete phase shift values is given as $\left\{0, \frac{2 \pi}{2^{b}}, \ldots, \frac{\left(2^{b}-1\right) 2 \pi}{2^{b}}\right\}$. Simulation shows that although quantization over phase shifts degrades system performance, 2-bits quantification RIS reflection elements are sufficient to achieve a large portion of ESE for the CPSs setup.

\begin{figure}[!htbp]
\centering
\includegraphics[width=3.5in]{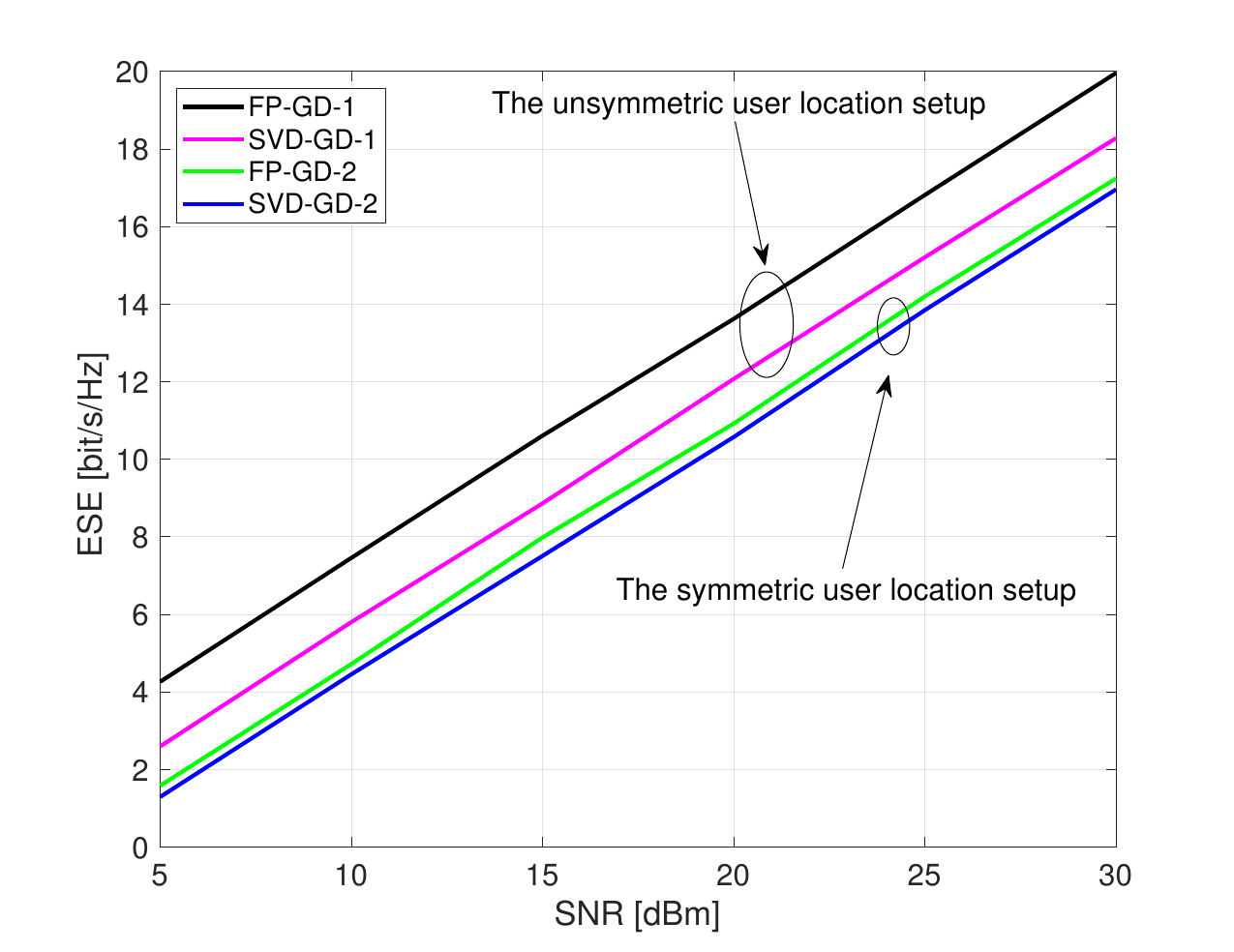}
\caption{AP transmit power versus ESE for different user setups.}\label{tu9}
\end{figure}

Fig.~\ref{tu9} illustrates the sum ESE versus AP transmit power for the user setup in Table~\ref{tabel1} and Table~\ref{tabel2}. It can be observed that the performance gap between the SVD-GD algorithm and the FP-GD algorithm is smaller under symmetric user location setup. In the SVD-GD algorithm, we assume the downlink beamforming vector for each user at the AP is equipped with the same transmit power. Since the utility function in (P1) is symmetric with respect to SINR, it is near optimal to allocate the same transmit power for all users when we take the symmetric user location setup in Table~\ref{tabel1}, where the channels are statistically equally strong. This is the reason why the performance degradation of the SVD-GD algorithm is smaller under symmetric user location setup.

\section{CONCLUSION}
\label{CONCLUSION}
\quad This paper studied beamforming design in RIS-aided MISO downlink system with correlated channels. We presented two joint beamforming algorithms, namely the SVD-GD algorithm and the FP-GD algorithm. For the SVD-GD algorithm, no alternative optimization is required for the joint beamforming, however its active beamforming strategy is suboptimal since the power is allocated evenly among all beamforming direction. For the FP-GD algorithm, its active beamforming vectors are designed more carefully, however alternative optimization is required for joint beamforming. Simulation results showed that the proposed algorithm achieves decent performance compared with benchmark algorithms. Besides, we confirmed that 2-bits quantification RIS is sufficient to achieve a large portion of the spectral efficiency for the CPSs setup.

\section*{ACKNOWLEDGEMENT}
\label{ACKNOWLEDGEMENT}
\quad This work is partially supported by the National Key Research and Development Project under Grant 2020YFB1806805 and Science and Technology on Communication Networks Laboratory. The work of Haochen Li was supported by China
Scholarship Council.

\bibliographystyle{IEEEtran}
\bibliography{myref}



\end{document}